\documentclass[conference]{IEEEtran}
\IEEEoverridecommandlockouts

\usepackage{graphicx}
\usepackage{textcomp}
\usepackage{hyperref,url}

\usepackage{graphicx, xcolor}

\usepackage{adjustbox}

\usepackage{enumitem}
  \setlist{topsep=2pt,partopsep=2pt,noitemsep}

\usepackage{amsmath, amsfonts, amssymb}

\usepackage[algoruled, linesnumbered]{algorithm2e}

\usepackage{tikz, pgfplots}
\pgfplotsset{compat=newest}

\DeclareMathOperator*{\argmin}{argmin}

\newcommand{\qm}[1]{``#1''}

\newcommand{\x}{\mathbf{x}} 
\newcommand{\xorig}{\mathbf{x}^{\textup{orig}}} 
\newcommand{\noise}{\mathbf{n}} 
\newcommand{\Noise}{\mathbf{N}} 
\newcommand{\z}{\mathbf{z}} 
\newcommand{\dual}{\mathbf{u}} 
\newcommand{\coef}{\mathbf{a}} 
\newcommand{\e}{\mathbf{e}} 
\newcommand{\mask}{\mathbf{m}} 
\newcommand{\xcor}{\mathbf{x}^{\textup{cor}}} 

\newcommand{\X}{\mathbf{X}} 
\newcommand{\Xorig}{\mathbf{X}^{\textup{orig}}} 
\newcommand{\Xcor}{\mathbf{X}^{\textup{cor}}} 
\newcommand{\Tfmask}{\mathbf{M}} 
\newcommand{\A}{\mathbf{A}} 
\newcommand{\ana}{\mathbf{L}} 
\newcommand{\Id}{\mathbf{I}} 

\newcommand{\Cset}{\mathbb{C}} 
\newcommand{\Rset}{\mathbb{R}} 
\newcommand{\ind}[1]{\iota_{#1}} 
\newcommand{\sigset}{\mathit{\Gamma}_{\textup{T}}} 
\newcommand{\coefset}{\mathit{\Gamma}_{\textup{TF}}} 

\newcommand{\norm}[1]{\|#1\|}
\newcommand{\proj}[1]{\operatorname{proj}_{#1}}
\newcommand{\prox}[1]{\operatorname{prox}_{#1}}

\title{%
Janssen 2.0: Audio Inpainting in the Time-frequency Domain
\thanks{The work was supported by the Czech Science Foundation (GA\v{C}R) Project No.\,23-07294S.
The authors are grateful to NVIDIA for their donation of the Titan~XP graphic card.}
}

\author{%
\IEEEauthorblockN{Ondřej Mokrý}
\IEEEauthorblockA{\textit{Dept.\ of Telecommunications} \\
\textit{Brno University of Technology}\\
Czech Republic\\
\href{mailto:ondrej.mokry@vut.cz}{ondrej.mokry@vut.cz}}
\and
\IEEEauthorblockN{Peter Balušík}
\IEEEauthorblockA{\textit{Dept.\ of Telecommunications} \\
\textit{Brno University of Technology}\\
Czech Republic\\
\href{mailto:230531@vut.cz}{230531@vut.cz}}
\and
\IEEEauthorblockN{Pavel Rajmic}
\IEEEauthorblockA{\textit{Dept.\ of Telecommunications} \\
\textit{Brno University of Technology}\\
Czech Republic\\
\href{mailto:pavel.rajmic@vut.cz}{pavel.rajmic@vut.cz}}
}

\begin{document}
%
\maketitle
\begin{abstract}
The paper focuses on inpainting missing parts of an audio signal spectrogram, i.e., estimating the lacking time-frequency coefficients.
The autoregression-based Janssen algorithm, a state-of-the-art for the time-domain audio inpainting,
is adapted for the time-frequency setting.
This novel method, termed Janssen-TF, is compared with the deep-prior neural network approach using both objective metrics and a~subjective listening test,
proving Janssen-TF to be superior in all the considered measures.
\end{abstract}
\begin{IEEEkeywords}
    audio inpainting, autoregression, deep prior, DPAI, time-frequency, spectrogram.
\end{IEEEkeywords}

\section{Introduction}

Audio inpainting, sometimes referred to as missing audio interpolation or concealment,
is the task of filling in the absent parts of audio with a~meaningful content.
Typically, the missing information is in the form of lost samples in the \textit{time domain}.
If compact regions of such samples are present, they are often referred to as the gaps.
For the described task, numerous methods have been developed:
those exploiting the autoregressive character of audio
\cite{Etter1996:Interpolation_AR,
javevr86},
methods based on the sparsity of the time-frequency representation of audio
\cite{Adler2012:Audio.inpainting,
MokryRajmic2020:Inpainting.revisited,
MokryZaviskaRajmicVesely2019:SPAIN,
Mokry2022:Audio.inpainting.NMF},
or methods relying on similarities in audio recordings
\cite{Bahat_2015:Self.content.based.audio.inpaint,
ToumiEmiya2018:Sparse.Non.Local.Inpainting,
Perraudin2018:Similarity.Graphs}.
Quite recently, deep learning audio inpainting methods appeared
\cite{Greshler2021:Catch.A.Waveform,
Moliner2023:Audio.inverse.problems.diffusion}.
For gaps of up to 50~ms in length, the Janssen method
\cite{javevr86,MokryRajmic2025:Inpainting.AR} has been evaluated the best for a~long time,
but PHAIN \cite{TanakaYatabeOikawa2024:PHAIN} may become a~new state-of-the art.

The time-domain inpainting represents the most natural case from the point of view of the application to signal recovery.
Nonetheless, many practical algorithms,
such as classic audio codecs
(MP3, AAC, etc.),
operate on short-time signal spectra.
From this perspective, it is valuable to think about inpainting in the \textit{time-frequency domain},%
\footnote{In other fields, one can go even further and require inpainting purely in the frequency domain, see for example \cite{DankovaRajmicJirik2015:LVA},
\cite{RajmicKoldovskyDankova2017:Fast.reconstruction.sparse.relative.impulse.responses}.%
}
besides the fact that it is an interesting problem per se.
In the case under consideration, a~portion of time-frequency (TF) coefficients%
\footnote{
typically in the form of a complex-valued matrix
}
is missing, which need to be recovered to allow a~later reconstruction of the time-domain signal.
Indeed, there have been a~few methods developed in the
TF context, all of them relying on neural modeling
\cite{Marafioti2019:Context.encoder,
Marafioti2021:GACELA,%
Miotello2023:Deep.Prior.Inpainting.Harmonic.CNN}.
The most recent work
\cite{Miotello2023:Deep.Prior.Inpainting.Harmonic.CNN}
is based on the concept of a~\qm{deep prior}\!,
a~neural network that does not need training data besides the corrupted observation
\cite{UlyanovVedaldiLempitsky2020:Deep.Image.Prior-journal}. 


The goal of this paper is to adapt the Janssen algorithm~\cite{javevr86},
proven successful in the time-domain case, for the TF-domain case,
and compare its performance with the recent deep-prior-based approach of
\cite{Miotello2023:Deep.Prior.Inpainting.Harmonic.CNN},
both using objective metrics and by listening tests.


\section{Gaps and the Short-time Fourier Transform}
\label{Sec:About.gaps.and.STFT}
Note that a~compact gap in the time domain is far from being equivalent to a~gap in the TF 
domain, which can be associated with the uncertainty principle.
When a~time-domain signal is transformed using the Short-time Fourier Transform (STFT),
it is first divided into overlapping frames, which are weighted using windows such as the Hann window (see, e.g., \cite{badokoto13}).
When a~gap is present, the number of TF coefficients affected by the gap depends on how much the frames overlap.
In practical scenarios, the gap overlaps with more than one frame;
setting the missing samples to zeros then reduces the value of the corresponding TF coefficients, leading to a~kind of TF-fadeout.
For wide gaps,
even a~number of completely zero columns can appear in the TF domain (i.e., in the spectrogram),
see the analysis in \cite{RajmicBartlovaPrusaHolighaus2015:Acceleration.support.restriction}.
Since the inverse STFT also uses windowed frames, the same holds in the opposite direction:
consecutive empty spectrogram columns result in a decrease in the time-domain amplitude, governed by the window shape.
In turn, there is no way how to equitably compare time-domain inpainting techniques with those specialized for the TF-domain.


\section{Deep Prior Audio Inpainting (DPAI)}

The
deep prior, introduced in \cite{UlyanovVedaldiLempitsky2020:Deep.Image.Prior-journal},
is a~concept used for the reconstruction of corrupted signals using
a~convolutional neural network (CNN).
In contrast to most CNN applications,
\cite{UlyanovVedaldiLempitsky2020:Deep.Image.Prior-journal}
does not utilize a~training dataset;
the reconstruction only involves a~proper CNN architecture and the corrupted signal.
The deep prior concept has also proven that the network architecture matters significantly more than previously thought.

\subsection{Application of Deep Prior to Audio Inpainting}
\label{ssec:dpai}
The deep prior reconstruction can be utilized in a~variety of applications,
as outlined in
\cite{UlyanovVedaldiLempitsky2020:Deep.Image.Prior-journal}.
In the context of the time-domain audio inpainting, consider a corrupted signal
\mbox{\(\xcor\!\in\Rset^N\)}\!,
having missing regions (filled with zeros), compared with the original 
signal \(\xorig\!\in\Rset^N\)\!.
Formally, if a~binary mask
\(\mask\in \{0,1\}^{N}\)
is utilized to represent the missing sections of the signal, the relation
$\xcor = \mask\odot\xorig$
holds, where \(\odot\) represents the elementwise product.
Such a relation
naturally characterizes the feasible set
$\sigset = \{ \x \,|\, \mask\odot \x = \mask\odot\xorig \}$.

An encoder--decoder CNN denoted 
$f_{\theta}(\noise)$ is assumed to be able to generate \emph{any} signal of size $N$ using a~random but fixed input noise
\(\noise\in\Rset^N\)\!,
using a~proper $\theta$.
This is a~reliable assumption since the number of network parameters \({\theta}\)
(weights and biases)
is significantly greater than $N$.

In the deep prior reconstruction, 
also the parameters $\theta$ are initialized randomly.
To obtain a~reconstruction result, \(\theta\) is then optimized as follows:
\begin{equation}
    \label{data_prior}
    \theta^* = \argmin_{\theta} E(\mask\odot f_{\theta}(\noise),\xcor),
\end{equation}
where
\(E\) is a~differentiable data fidelity function measuring the proximity of its two inputs.
The optimal \(\theta^*\) can be obtained using an iterative optimizer, e.g., Adam \cite{Kingma2014:Adam}.
The crucial observation is that the optimization must be interrupted before it reaches $\theta^*$\!,
for which we would have
$f_{\theta^*\!}(\noise) = \xcor$\!,
meaning basically overfitting.
Instead, early stopping of the training is desired,
resulting in an estimate $\hat\theta$ and a~corresponding $\hat{\x} = f_{\hat{\theta}}(\noise)$
\cite{UlyanovVedaldiLempitsky2020:Deep.Image.Prior-journal}.

The authors of \cite{Miotello2023:Deep.Prior.Inpainting.Harmonic.CNN}
used an analog approach to inpaint audio in the 
time-frequency
domain.
Through the use of the STFT,
denoted $\ana$ in the following,
an audio recording \(\x\) is considered merely in the TF domain, where it forms a~spectrogram matrix
\(\ana\x = \X \in\Cset^{T\times F}\)\!.
Therefore the CNN operates on spectrograms and 
$\hat{\X} = f_{\hat{\theta}}(\Noise)$ for a~random but fixed matrix $\Noise$. 
Once $\hat{\X}$ has been established, the time-domain solution is obtained by the inverse STFT.

Regarding the mask, in \cite{Miotello2023:Deep.Prior.Inpainting.Harmonic.CNN}
only the case of completely missing columns of the spectrogram is considered,
i.e., for the TF mask
$\Tfmask\in\{0,1\}^{T\times F}$\!,
each of its columns is either full of ones or zeros.
Nevertheless, a~generalization to other TF masks is straightforward.
The observed spectrogram \(\Xcor\) is finally modeled
as
\( \Xcor\! = \Tfmask\odot\Xorig\),
where $\Xorig$ corresponds to $\xorig$ in the time domain,
i.e., $\Xorig = \ana\xorig$\!.
Note that by analogy
with the time-domain case,
we obtain the feasible set
$\coefset = \{ \X \,|\, \Tfmask\odot \X = \Tfmask\odot\Xorig \}$.

In \cite{Miotello2023:Deep.Prior.Inpainting.Harmonic.CNN},
the loss function $E$
in \eqref{data_prior} linearly combines the classic Mean Squared Error between spectrograms and
the multi-scale spectrogram loss
\cite{Yamamoto2020:Parallel.waveGAN.multi-resolution.sgram,Moliner2025:Diffusion.Adversarial.Nonlinear.Effects}.

\subsection{Architecture}
The authors of \cite{Miotello2023:Deep.Prior.Inpainting.Harmonic.CNN}
used an architecture similar to U-Net called MultiResUNet \cite{Ibtehaz2020:MultiResUNet}.
It is a common CNN architecture with several layers;
the skip connections between the encoder and decoder stages play a significant role in audio inpainting.
The architecture additionally includes the so-called harmonic convolution \cite{Takeuchi2020:HarmonicLowering},
which proved to be the best among multiple tested architectures in  
\cite{Miotello2023:Deep.Prior.Inpainting.Harmonic.CNN}.
The harmonic convolution is a~variation on the common 2D convolution that takes into account higher harmonics in audio.

  

\subsection{Variants}
The DPAI intrinsically reconstructs the whole $\hat\X$.
Yet, considering that only 
a~part of the spectrogram was not observed, 
the authors of 
\cite{Miotello2023:Deep.Prior.Inpainting.Harmonic.CNN}
evaluate the performance of their method after injecting the reliable coefficients back to the solution as a~postprocessing step.
Formally, such an action is a~projection
\begin{equation}
    \label{eq:projcoef}
    \proj{\coefset}(\hat\X) = \Tfmask\odot\Xcor + (\mathbf{1}-\Tfmask)\odot\hat{\X}
\end{equation}
onto the set $\coefset$.
Note that a resulting sharp transition between
the regions is smeared in the time domain due to the nature of the STFT
\cite{RajmicBartlovaPrusaHolighaus2015:Acceleration.support.restriction}; see also Sec.\ \ref{Sec:About.gaps.and.STFT}.
We call the reconstructions modified by \eqref{eq:projcoef} \emph{with context}, while the others using the entire $\hat\X$ are \emph{without context}.
%
%
%
In the experimental part, we will study whether employing the context brings
an improvement in performance.



\section{Janssen spectrogram inpainting (Janssen-TF)}


A signal $\x = [x_1,\dots, x_N]^\top\!\in\Rset^N$ corresponds to 
an autoregressive process if 
the following holds:
%
\begin{equation}
    \sum_{i=1}^{p+1} a_{i} x_{n+1-i} = e_{n}, \quad n = 1,\dots, N+p.
    \label{eq:error.sum}
\end{equation}
Here, $\coef = [1, a_2, \dots, a_{p+1}]^\top$ is the vector of model parameters (AR~coefficients), $p$ is the model order,
and $\x$ is zero-padded such that its final length is $N+p$.
The error term $\e\in\Rset^{N+p}$ represents a realization of a zero-mean white noise process with variance $\sigma^2$ \cite[Def.\,3.1.2]{BrockwellDavis2006:Time.series}.

In the \emph{time-domain} inpainting method proposed by Janssen et al.\ \cite{javevr86},
the restored audio signal is supposed to follow the AR model \eqref{eq:error.sum}
while both the missing samples in the time-domain and the unknown AR
parameters $\coef$ are being estimated. 
Janssen et al.\ approach the problem iteratively
by performing two principal steps in the $i$-th iteration:
\begin{enumerate}
    \item
    \label{itm:coef}
    for a temporary solution $\x^{(i)}$\!, estimate the AR model parameters $\coef^{(i)}$;
    \item
    \label{itm:signal}
    given the model parameters $\coef^{(i)}$\!,
    estimate a new temporary solution $\x^{(i+1)}$ which fits the observed samples and the norm of the model error $\e$ is minimized.
\end{enumerate}
Step \ref{itm:coef} can be solved efficiently using standard methods, such as the Durbin--Levinson recursion (see for example the \href{https://www.mathworks.com/help/signal/ref/lpc.html}{\texttt{lpc}} function in Matlab).
Step \ref{itm:signal} boils down to a mean squares problem which can be treated e.g.\ via matrix inversion.

The proposed adaptation of the Janssen iteration to the TF setting, denoted Janssen-TF, lies in modifying step \ref{itm:signal}.
In our case,
the \emph{signal spectrogram} is constrained to fit the observation, $\Xcor$\!,
at the reliable positions,
while the \emph{signal itself} follows the estimated AR model.
To formalize this modification, 
note that the error in \eqref{eq:error.sum} can be written as $\e = \A\x$,
where $\A\in\Rset^{(N+p)\times N}$ is a~Toeplitz matrix composed using the (fixed) coefficients $\coef$.
We can then pose the proposed modification of step \ref{itm:signal} as
\begin{equation}
    \x^{(i+1)} = \arg\min_{\x} \tfrac{1}{2}\norm{\A\x}^2 + \ind{\coefset}(\ana\x),
    \label{eq:subproblem}
\end{equation}
where
$\ana$ is the STFT
and
$\ind{\coefset}$ denotes the indicator function of the convex set $\coefset$ of feasible spectrograms.%
\footnote{%
    The original Janssen algorithm minimizes the function $\frac{1}{2}\norm{\A\x} + \ind{\sigset}(\x)$
    in subproblem \ref{itm:signal}.
}
As such, the problem \eqref{eq:subproblem} is convex and it is solvable using standard numerical methods.
We propose to employ the alternating direction method of multipliers (ADMM),
which includes two principal steps \cite[Sec.\,3.1.1]{Boyd2011ADMM}:
\begin{subequations}
    \begin{align}
    \bar\x^{(k+1)} &= \arg\min_{\x} \tfrac{1}{2}\norm{\A\x}^2 + \tfrac{\rho}{2}\norm{\ana\x - \z^{(k)} + \dual^{(k)}}^2\!,
    \label{eq:admm:x} \\
    \z^{(k+1)} &= \arg\min_{\z} \ind{\coefset}(\z) + \tfrac{\rho}{2}\norm{\ana\bar\x^{(k+1)} - \z + \dual^{(k)}}^2\!. \hspace{-1em}
    \label{eq:admm:z} 
\end{align}
\end{subequations}
The whole algorithm is summarized in Alg.\,\ref{alg:janssen.tf.admm}.
For the simplification of \eqref{eq:admm:x},
we have posed the assumption%
\footnote{%
    This assumption is met for instance if $\ana$ corresponds to a Parseval tight frame, which can be ensured by a~suitable practical setting of the STFT.}
$\ana^*\ana=\Id$,
leading to $\bar\x^{(k+1)} = \prox{\frac{1}{2}\norm{\cdot}^2}(\ana^*(\z^{(k)}-\dual^{(k)}))$,
see \cite[Rem.\,3]{ZaviskaMokryRajmic2018:SPADE_DetailedStudy},
which then corresponds to line \ref{ln:admm.x} of Alg.\ \ref{alg:janssen.tf.admm} \cite[Table~1]{combettes2011proximal}.
Similarly, \eqref{eq:admm:z}
boils down 
to line \ref{ln:admm.z} of the algorithm;
this is computed via \eqref{eq:projcoef}.
The hyperparameters of the algorithm are the number of outer and inner iterations ($I$ and $K$, respectively), and the step size~$\rho > 0$.

\begin{algorithm}
    \caption{Janssen-TF-ADMM}
    \label{alg:janssen.tf.admm}
    \small
    initialize $\x^{(1)}$ \\
    \For{$i=1,\dots,I$}{
        \tcp{AR model update}
        $\coef^{(i)} = \texttt{lpc}(\x^{(i)})$ \\
        \tcp{signal update using ADMM}
        form the Toeplitz matrix $\A$ from $\coef^{(i)}$ \\
        $\z^{(1)} = \ana\x^{(i)}$, $\dual^{(1)} = \mathbf{0}$  \\
        \For{$k=1, \dots, K$}{
            $\bar\x^{(k+1)} = (\Id + \frac{1}{\rho}\A\!^\top\!\A)^{-1}\ana^*(\z^{(k)} - \dual^{(k)})$
            \label{ln:admm.x} \\
            $\z^{(k+1)} = \proj{\coefset}(\ana\bar\x^{(k+1)} + \dual^{(k)})$
            \label{ln:admm.z} \\
            $\dual^{(k+1)} = \dual^{(k)} + \ana\bar\x^{(k+1)} - \z^{(k+1)}$
        }
        $\x^{(i+1)} = \bar\x^{(K+1)}$ \\
    }
\end{algorithm}

Note that several other algorithms can be used to solve \eqref{eq:subproblem} numerically,
such as the primal--dual algorithm \cite{ChambollePock2011:First-Order.Primal-Dual.Algorithm}
or the Douglas--Rachford or the forward--backward algorithms \cite{combettes2011proximal}, combined with the approximal operator of the function $\ind{\coefset}(\ana\,\cdot)$ \cite{MokryRajmic2020:Approximal.operator}.
However,
the variant with ADMM described above,
denoted Janssen-TF-ADMM, exhibited the best performance out of six algorithms in our preliminary testing.

\section{Experiments}

\subsection{Data, setup and metrics}
For our first experiment, we adopt the audio dataset%
\footnote{\url{https://github.com/fmiotello/dpai/}}
used for the evaluation of DPAI \cite{Miotello2023:Deep.Prior.Inpainting.Harmonic.CNN}.
It consists of 6~music and 2~speech recordings of 5~seconds in length, sampled at 16\,kHz.
Only minor modifications were made to the original distribution of gaps
(i.e., the placement of the missing spectrogram columns),
such that each test excerpt includes 5~gaps of the same length.
The gap lengths range from 1~to~6 missing columns.
In turn, the entire test set comprises $8\times6=48$~corrupt spectrograms.
Even though the dataset is relatively small and includes only a~few types of audio signals,
it provides a~fair comparison of the proposed method with DPAI---%
the choice allows the use of exactly the same 
U-net architecture 
of DPAI from \cite{Miotello2023:Deep.Prior.Inpainting.Harmonic.CNN}.
In particular, we chose their \qm{best2} setting,
and 5000 epochs of the Adam optimizer.

To validate the results on a larger dataset,
we hand-picked 60 music recordings of various complexity from the IRMAS dataset \cite{Bosch2012:IRMAS.paper,Bosch2014:IRMAS.dataset}.
To enable the use of DPAI, the recordings were subsampled to 16\,kHz and shortened to 5~seconds.

The STFT (i.e., the operator $\ana$) utilizes a~2048-sample-long Hann window
(corresponding to 128\,ms\footnote{This exceeds the standard window length for speech processing; however, a~change of this hyperparameter would cause the need for changing the DPAI architecture.}), 
with 75\% overlap and 2048 frequency channels; this setting is the default for the time-frequency transform of the librosa package\footnote{\url{https://librosa.org/doc/0.10.2/generated/librosa.stft.html}}\!,
as used in
\cite{Miotello2023:Deep.Prior.Inpainting.Harmonic.CNN},
and ensures that $\ana$ forms a~Parseval tight frame.

We compare the proposed method Janssen-TF-ADMM with three baseline methods:
\begin{itemize}
    \item DPAI reconstruction with context,
    \item DPAI reconstruction without context,
    \item gap-wise Janssen method in the time domain \cite{MokryRajmic2025:Inpainting.AR}.
\end{itemize}
The gap-wise method is a~recent twist to the original Janssen algorithm \cite{javevr86},
which is shown in \cite{MokryRajmic2025:Inpainting.AR} to be the state-of-the-art time-domain inpainter.
As explained in Sec.~\ref{Sec:About.gaps.and.STFT}, the time-domain Janssen method is not directly comparable with the first two.
We use it such that we transform the corrupted spectrogram \(\Xcor\) to the time-domain and zero-out all samples affected by the zero columns of \(\Xcor\).

To objectively evaluate the reconstruction quality,
we assess the difference between the inpainted signal $\hat\x$ and the clean reference $\xorig$ using two standard metrics:
\begin{enumerate}
    \item signal-to-noise ratio (SNR), defined in decibels as
    $\text{SNR}(\hat\x, \xorig) = 10\log_{10}\frac{\norm{\xorig}^2}{\norm{\xorig-\hat\x}^2}$ \cite{Adler2012:Audio.inpainting},
    \item objective difference grade (ODG) computed using the PEMO-Q \cite{Huber:2006a},
    which takes psychoacoustics into account and predicts the subjective evaluation of the difference between $\hat\x$ and $\xorig$
    on the scale from $-4$ to~$0$
    (from \qm{very annoying} to \qm{imperceptible}).
\end{enumerate}
In addition, we run
\begin{enumerate}
    \setcounter{enumi}{2}
    \item 
    a~subjective listening test containing a~limited number of excerpts.
    The MUSHRA-type test
    \cite{ITU-R2015:MUSHRA}
    was performed, which includes
    the reference signal,
    the competing recosntructions,
    and finally the hidden anchor, i.e., the signal reconstructed from the spectrogram with gaps.
    The participants were asked to score the perceived similarity of the signals with the reference signal on a~continuous scale from 0 to 100,
    with the help of the webMUSHRA environment%
    \footnote{\url{https://github.com/audiolabs/webMUSHRA/}}\!.
    The test was run in a~quiet music studio, using a~professional sound card and headphones.
    The listening conditions were identical for all the participants.
    Following the recommendation \cite{ITU-R2015:MUSHRA},
    11 assessors out of 18 were selected through 
    post-screening.
\end{enumerate}
%

\subsection{Results}




The objective comparison on the DPAI dataset 
is shown in Fig.\ \ref{fig:lineplot}. 
The first observation is that the time-domain-based Janssen method 
\cite{MokryRajmic2025:Inpainting.AR}
is not competitive with the other approaches.
While the average performance of the DPAI variants mostly lies within the confidence intervals of each other,
suggesting limited significance of the difference,
the proposed Janssen-TF-ADMM method largely outperforms the competitors in both SNR and ODG
(except the longest gaps).
%

\begin{figure}[t]
    \centering
    %
    %
    \adjustbox{width=0.72\linewidth}{
%
%
\definecolor{mycolor1}{rgb}{0.00000,0.44700,0.74100}%
\definecolor{mycolor2}{rgb}{0.85000,0.32500,0.09800}%
\definecolor{mycolor4}{rgb}{0.92900,0.69400,0.12500}%
\definecolor{mycolor5}{rgb}{0.49400,0.18400,0.55600}%
\begin{tikzpicture}[trim axis left, trim axis right]

\begin{axis}[%
width=4.00in,
height=2.55in,
at={(0.758in,0.481in)},
scale only axis,
xmin=1,
xmax=6,
xtick={1, 2, 3, 4, 5, 6},
xlabel style={text=white!15!black},
xlabel={gap length},
ymin=0,
ymax=60,
ylabel style={text=white!15!black},
ylabel={SNR (dB)},
axis background/.style={fill=white},
xmajorgrids,
ymajorgrids,
legend style={legend cell align=left, align=left, draw=white!15!black}
]

\addplot[area legend, draw=black, fill=mycolor1, draw opacity=0.1, fill opacity=0.1, forget plot]
table[row sep=crcr] {%
x	y\\
1	16.4114131823584\\
2	17.4672823567892\\
3	13.0004701313859\\
4	9.93340991999562\\
5	8.73991313719166\\
6	7.35750371511233\\
6	11.2021777936956\\
5	11.9476678412599\\
4	13.5089961709654\\
3	18.0949132328418\\
2	29.2771114529016\\
1	24.9579755436761\\
}--cycle;
\addplot [color=mycolor1, line width=1.0pt, dashed]
  table[row sep=crcr]{%
1	19.8488077521324\\
2	20.6064563989639\\
3	14.9709348380565\\
4	11.2200104445219\\
5	10.4656670987606\\
6	9.34696659445761\\
};
\addlegendentry{DPAI without context}

\addplot[area legend, draw=black, fill=mycolor2, draw opacity=0.1, fill opacity=0.1, forget plot]
table[row sep=crcr] {%
x	y\\
1	23.7566328002682\\
2	18.4160402693953\\
3	13.2080473295931\\
4	10.3817894265965\\
5	9.03730023193692\\
6	7.67113015688136\\
6	11.5099985274897\\
5	12.3494729042637\\
4	13.9924696487416\\
3	18.4818137089173\\
2	27.6488478220931\\
1	33.7669188027157\\
}--cycle;
\addplot [color=mycolor2, line width=1.0pt, dotted]
  table[row sep=crcr]{%
1	27.8477057814597\\
2	22.8286235034465\\
3	15.5029974877834\\
4	11.5515997260808\\
5	10.928790718317\\
6	9.46995750069616\\
};
\addlegendentry{DPAI with context}


\addplot[area legend, draw=black, fill=mycolor4, draw opacity=0.1, fill opacity=0.1, forget plot]
table[row sep=crcr] {%
x	y\\
1	8.36541476507532\\
2	7.70620286605013\\
3	6.98185837829857\\
4	5.53462143195627\\
5	6.07240032861253\\
6	4.57885074413695\\
6	8.30707103395538\\
5	10.2470903415551\\
4	9.24108976771823\\
3	11.1496268401962\\
2	13.6221236770118\\
1	15.4248720281268\\
}--cycle;
\addplot [color=mycolor4, line width=1.0pt, dashdotted]
  table[row sep=crcr]{%
1	11.283407834\\
2	10.269642038375\\
3	8.65852969325\\
4	7.522729536\\
5	7.521091364125\\
6	6.638937442375\\
};
\addlegendentry{gap-wise Janssen}

\addplot[area legend, draw=black, fill=mycolor5, draw opacity=0.1, fill opacity=0.1, forget plot]
table[row sep=crcr] {%
x	y\\
1	60.7208209104834\\
2	50.1804241890775\\
3	21.2856081139968\\
4	14.4410005159851\\
5	11.6096602014468\\
6	8.89684471407023\\
6	15.6783812667734\\
5	18.5366572051725\\
4	23.676808818071\\
3	34.3754630575005\\
2	69.1266324075459\\
1	94.2791084887949\\
}--cycle;
\addplot [color=mycolor5, line width=1.0pt]
  table[row sep=crcr]{%
1	67.4841026075\\
2	57.52817312375\\
3	27.8814609425\\
4	17.51561540625\\
5	15.10456597225\\
6	12.122108814875\\
};
\addlegendentry{Janssen-TF-ADMM}

\end{axis}
\end{tikzpicture}%
}
    \adjustbox{width=0.72\linewidth}{
%
%
\definecolor{mycolor1}{rgb}{0.00000,0.44700,0.74100}%
\definecolor{mycolor2}{rgb}{0.85000,0.32500,0.09800}%
\definecolor{mycolor4}{rgb}{0.92900,0.69400,0.12500}%
\definecolor{mycolor5}{rgb}{0.49400,0.18400,0.55600}%
\begin{tikzpicture}[trim axis left, trim axis right]

\begin{axis}[%
width=4.00in,
height=2.55in,
at={(0.758in,0.481in)},
scale only axis,
xmin=1,
xmax=6,
xtick={1, 2, 3, 4, 5, 6},
xlabel style={text=white!15!black},
xlabel={gap length},
ymin=-3.5,
ymax=0,
ylabel style={text=white!15!black},
ylabel={ODG},
axis background/.style={fill=white},
xmajorgrids,
ymajorgrids,
legend style={legend cell align=left, align=left, draw=white!15!black}
]

\addplot[area legend, draw=black, fill=mycolor1, draw opacity=0.1, fill opacity=0.1, forget plot]
table[row sep=crcr] {%
x	y\\
1	-2.80525100397144\\
2	-2.36611301512488\\
3	-2.83711342623981\\
4	-2.99372762743945\\
5	-3.21065808176392\\
6	-3.09672973651899\\
6	-2.73937588864965\\
5	-2.70654874113239\\
4	-2.20584551217157\\
3	-1.33470649991753\\
2	-0.557414441666776\\
1	-1.05040090297938\\
}--cycle;
\addplot [color=mycolor1, line width=1.0pt, dashed]
  table[row sep=crcr]{%
1	-2.12311534699022\\
2	-1.54337055793337\\
3	-2.08217518557244\\
4	-2.77874244375287\\
5	-3.01534655051947\\
6	-2.9956422165679\\
};
\addlegendentry{DPAI without context}

\addplot[area legend, draw=black, fill=mycolor2, draw opacity=0.1, fill opacity=0.1, forget plot]
table[row sep=crcr] {%
x	y\\
1	-1.09604386174508\\
2	-1.73175114013617\\
3	-2.44173130321666\\
4	-2.8444364044739\\
5	-3.13489129285861\\
6	-3.08487226859027\\
6	-2.68223052809542\\
5	-2.62875626086734\\
4	-2.2954261473396\\
3	-1.46263064420178\\
2	-0.550602642786058\\
1	-0.185012481353407\\
}--cycle;
\addplot [color=mycolor2, line width=1.0pt, dotted]
  table[row sep=crcr]{%
1	-0.380283486648846\\
2	-0.890401349772898\\
3	-1.92058866767568\\
4	-2.62233487796798\\
5	-2.94997132465317\\
6	-2.94126618608791\\
};
\addlegendentry{DPAI with context}


\addplot[area legend, draw=black, fill=mycolor4, draw opacity=0.1, fill opacity=0.1, forget plot]
table[row sep=crcr] {%
x	y\\
1	-3.19331455267432\\
2	-3.23618331858604\\
3	-3.30144892907346\\
4	-3.2906570231989\\
5	-3.31210248444025\\
6	-3.31890068790721\\
6	-3.14299320266096\\
5	-2.86756492046204\\
4	-2.97587111004098\\
3	-2.99059589200806\\
2	-2.96892150579037\\
1	-1.57494791560981\\
}--cycle;
\addplot [color=mycolor4, line width=1.0pt, dashdotted]
  table[row sep=crcr]{%
1	-2.9613055395\\
2	-3.11217998075\\
3	-3.17229600325\\
4	-3.15591072275\\
5	-3.231922756\\
6	-3.245245389625\\
};
\addlegendentry{gap-wise Janssen}

\addplot[area legend, draw=black, fill=mycolor5, draw opacity=0.1, fill opacity=0.1, forget plot]
table[row sep=crcr] {%
x	y\\
1	-0.00637343379358408\\
2	-0.276271892078195\\
3	-1.47313605717119\\
4	-2.56297820556221\\
5	-3.00656977806505\\
6	-3.11818825373146\\
6	-2.63141236381455\\
5	-1.96735709743695\\
4	-1.11101970767873\\
3	-0.237197922180698\\
2	-0.00207386314486149\\
1	-0.000875473601390054\\
}--cycle;
\addplot [color=mycolor5, line width=1.0pt]
  table[row sep=crcr]{%
1	-0.0033654339625\\
2	-0.032930849375\\
3	-0.584994203125\\
4	-1.879562247125\\
5	-2.667150055375\\
6	-2.9510586395\\
};
\addlegendentry{Janssen-TF-ADMM}

\end{axis}
\end{tikzpicture}%
}%
    %
    %
    %
    \vspace{-7pt}
    \caption{%
        Comparison of the inpainting methods using
        objective metrics -- SNR (top) and ODG (bottom).
        The plots show results averaged over the 8~test signals
        together with the bootstrap interval estimates of the mean values at the $\alpha=5\%$ significance level
        \cite[p.\,160]{EfronTibshirani:Bootstrap}.
        If the intervals do not overlap, it may be concluded that the difference of the means is statistically significant. 
    }
    \label{fig:lineplot}
\end{figure}

To support the objective evaluation on the first dataset,
Fig.~\ref{fig:boxplot} presents the results of the subjective listening test.
Its results correlate well with the objective evaluation;
in particular,
Janssen-TF-ADMM is superior to the other methods.
Moreover, this result is statistically significant in terms of medians.
The DPAI variants scored almost the same.

\begin{figure}[t]
    \vspace{4pt}
    \centering
    \adjustbox{width=.95\linewidth}{
%
%
\definecolor{mycolor1}{rgb}{0.75000,0.86175,0.93525}%
\definecolor{mycolor2}{rgb}{0.00000,0.44700,0.74100}%
\definecolor{mycolor3}{rgb}{0.85000,0.32500,0.09800}%
\begin{tikzpicture}

\begin{axis}[%
width=4.00in,
height=2.500in,
at={(0.758in,0.755in)},
scale only axis,
unbounded coords=jump,
xmin=0.5,
xmax=6.5,
xtick={1,2,3,4,5,6},
xticklabels={{anchor},{reference},{DPAI \\ without context},{DPAI \\ with context},{gap-wise \\ Janssen},{Janssen-\\-TF-ADMM}},
xticklabel style={rotate=45, align=right},
ymin=-5,
ymax=105,
ylabel style={font=\color{white!15!black}},
ylabel={MUSHRA score},
axis background/.style={fill=white},
title style={font=\bfseries},
ymajorgrids,
yminorgrids,
major grid style={white!75!black},
minor grid style={white!90!black},
minor y tick num=3,
]

\addplot[area legend, draw=none, fill=mycolor1, forget plot]
table[row sep=crcr] {%
x	y\\
0.875	0\\
0.75	1.33890049871323\\
1.25	1.33890049871323\\
1.125	0\\
1.25	0\\
0.75	0\\
0.875	0\\
}--cycle;

\addplot[area legend, draw=none, fill=mycolor1, forget plot]
table[row sep=crcr] {%
x	y\\
5.875	84\\
5.75	88.351426620818\\
6.25	88.351426620818\\
6.125	84\\
6.25	79.648573379182\\
5.75	79.648573379182\\
5.875	84\\
}--cycle;

\addplot[area legend, draw=none, fill=mycolor1, forget plot]
table[row sep=crcr] {%
x	y\\
2.875	71\\
2.75	75.2398515792586\\
3.25	75.2398515792586\\
3.125	71\\
3.25	66.7601484207414\\
2.75	66.7601484207414\\
2.875	71\\
}--cycle;

\addplot[area legend, draw=none, fill=mycolor1, forget plot]
table[row sep=crcr] {%
x	y\\
3.875	74\\
3.75	78.2398515792586\\
4.25	78.2398515792586\\
4.125	74\\
4.25	69.7601484207414\\
3.75	69.7601484207414\\
3.875	74\\
}--cycle;


\addplot[area legend, draw=none, fill=mycolor1, forget plot]
table[row sep=crcr] {%
x	y\\
4.875	30\\
4.75	32.9009510805453\\
5.25	32.9009510805453\\
5.125	30\\
5.25	27.0990489194547\\
4.75	27.0990489194547\\
4.875	30\\
}--cycle;
\addplot [color=mycolor2, dashed, forget plot]
  table[row sep=crcr]{%
1	12\\
1	29\\
};
\addplot [color=mycolor2, dashed, forget plot]
  table[row sep=crcr]{%
2	100\\
2	100\\
};
\addplot [color=mycolor2, dashed, forget plot]
  table[row sep=crcr]{%
3	88\\
3	100\\
};
\addplot [color=mycolor2, dashed, forget plot]
  table[row sep=crcr]{%
4	90\\
4	100\\
};
\addplot [color=mycolor2, dashed, forget plot]
  table[row sep=crcr]{%
5	45\\
5	74\\
};
\addplot [color=mycolor2, dashed, forget plot]
  table[row sep=crcr]{%
6	100\\
6	100\\
};
\addplot [color=mycolor2, dashed, forget plot]
  table[row sep=crcr]{%
1	0\\
1	0\\
};
\addplot [color=mycolor2, dashed, forget plot]
  table[row sep=crcr]{%
2	100\\
2	100\\
};
\addplot [color=mycolor2, dashed, forget plot]
  table[row sep=crcr]{%
3	15\\
3	50\\
};
\addplot [color=mycolor2, dashed, forget plot]
  table[row sep=crcr]{%
4	12\\
4	52\\
};
\addplot [color=mycolor2, dashed, forget plot]
  table[row sep=crcr]{%
5	0\\
5	19\\
};
\addplot [color=mycolor2, dashed, forget plot]
  table[row sep=crcr]{%
6	11\\
6	61\\
};
\addplot [color=mycolor2, forget plot]
  table[row sep=crcr]{%
0.875	29\\
1.125	29\\
};
\addplot [color=mycolor2, forget plot]
  table[row sep=crcr]{%
1.875	100\\
2.125	100\\
};
\addplot [color=mycolor2, forget plot]
  table[row sep=crcr]{%
2.875	100\\
3.125	100\\
};
\addplot [color=mycolor2, forget plot]
  table[row sep=crcr]{%
3.875	100\\
4.125	100\\
};
\addplot [color=mycolor2, forget plot]
  table[row sep=crcr]{%
4.875	74\\
5.125	74\\
};
\addplot [color=mycolor2, forget plot]
  table[row sep=crcr]{%
5.875	100\\
6.125	100\\
};
\addplot [color=mycolor2, forget plot]
  table[row sep=crcr]{%
0.875	0\\
1.125	0\\
};
\addplot [color=mycolor2, forget plot]
  table[row sep=crcr]{%
1.875	100\\
2.125	100\\
};
\addplot [color=mycolor2, forget plot]
  table[row sep=crcr]{%
2.875	15\\
3.125	15\\
};
\addplot [color=mycolor2, forget plot]
  table[row sep=crcr]{%
3.875	12\\
4.125	12\\
};
\addplot [color=mycolor2, forget plot]
  table[row sep=crcr]{%
4.875	0\\
5.125	0\\
};
\addplot [color=mycolor2, forget plot]
  table[row sep=crcr]{%
5.875	11\\
6.125	11\\
};
\addplot [color=mycolor2, forget plot]
  table[row sep=crcr]{%
0.875	0\\
0.75	1.33890049871323\\
0.75	12\\
1.25	12\\
1.25	1.33890049871323\\
1.125	0\\
1.25	0\\
1.25	0\\
0.75	0\\
0.75	0\\
0.875	0\\
};
\addplot [color=mycolor2, forget plot]
  table[row sep=crcr]{%
1.875	100\\
1.75	100\\
1.75	100\\
2.25	100\\
2.25	100\\
2.125	100\\
2.25	100\\
2.25	100\\
1.75	100\\
1.75	100\\
1.875	100\\
};
\addplot [color=mycolor2, forget plot]
  table[row sep=crcr]{%
2.875	71\\
2.75	75.2398515792586\\
2.75	88\\
3.25	88\\
3.25	75.2398515792586\\
3.125	71\\
3.25	66.7601484207414\\
3.25	50\\
2.75	50\\
2.75	66.7601484207414\\
2.875	71\\
};
\addplot [color=mycolor2, forget plot]
  table[row sep=crcr]{%
3.875	74\\
3.75	78.2398515792586\\
3.75	90\\
4.25	90\\
4.25	78.2398515792586\\
4.125	74\\
4.25	69.7601484207414\\
4.25	52\\
3.75	52\\
3.75	69.7601484207414\\
3.875	74\\
};
\addplot [color=mycolor2, forget plot]
  table[row sep=crcr]{%
4.875	30\\
4.75	32.9009510805453\\
4.75	45\\
5.25	45\\
5.25	32.9009510805453\\
5.125	30\\
5.25	27.0990489194547\\
5.25	19\\
4.75	19\\
4.75	27.0990489194547\\
4.875	30\\
};
\addplot [color=mycolor2, forget plot]
  table[row sep=crcr]{%
5.875	84\\
5.75	88.351426620818\\
5.75	100\\
6.25	100\\
6.25	88.351426620818\\
6.125	84\\
6.25	79.648573379182\\
6.25	61\\
5.75	61\\
5.75	79.648573379182\\
5.875	84\\
};
\addplot [color=mycolor3, line width=1.0pt, forget plot]
  table[row sep=crcr]{%
0.875	0\\
1.125	0\\
};
\addplot [color=mycolor3, line width=1.0pt, forget plot]
  table[row sep=crcr]{%
1.875	100\\
2.125	100\\
};
\addplot [color=mycolor3, line width=1.0pt, forget plot]
  table[row sep=crcr]{%
2.875	71\\
3.125	71\\
};
\addplot [color=mycolor3, line width=1.0pt, forget plot]
  table[row sep=crcr]{%
3.875	74\\
4.125	74\\
};
\addplot [color=mycolor3, line width=1.0pt, forget plot]
  table[row sep=crcr]{%
4.875	30\\
5.125	30\\
};
\addplot [color=mycolor3, line width=1.0pt, forget plot]
  table[row sep=crcr]{%
5.875	84\\
6.125	84\\
};
\addplot [color=black, only marks, mark=+, mark options={solid, draw=mycolor3}, forget plot]
  table[row sep=crcr]{%
1	31\\
1	31\\
1	31\\
1	33\\
1	34\\
1	37\\
1	38\\
1	40\\
1	41\\
1	41\\
1	44\\
1	49\\
1	62\\
1	62\\
};
\addplot [color=black, only marks, mark=+, mark options={solid, draw=mycolor3}, forget plot]
  table[row sep=crcr]{%
2	71\\
2	76\\
2	81\\
2	85\\
2	87\\
2	88\\
2	89\\
2	89\\
2	89\\
2	89\\
2	90\\
2	91\\
2	91\\
2	93\\
2	93\\
2	94\\
2	94\\
2	95\\
2	95\\
2	95\\
2	98\\
};
\addplot [color=black, only marks, mark=+, mark options={solid, draw=mycolor3}, forget plot]
  table[row sep=crcr]{%
nan	nan\\
};
\addplot [color=black, only marks, mark=+, mark options={solid, draw=mycolor3}, forget plot]
  table[row sep=crcr]{%
nan	nan\\
};
\addplot [color=black, only marks, mark=+, mark options={solid, draw=mycolor3}, forget plot]
  table[row sep=crcr]{%
5	85\\
5	85\\
};
\addplot [color=black, only marks, mark=+, mark options={solid, draw=mycolor3}, forget plot]
  table[row sep=crcr]{%
nan	nan\\
};
\end{axis}

\end{tikzpicture}%
}
    \vspace{-1ex}%
	\caption{%
        A boxplot showing the distribution of scores in the listening test.
        The individual boxes span from the 25th to the 75th percentile of the recorded scores.
        The notches (filled areas) around the medians (orange lines) are constructed such that boxes whose notches do not overlap have different medians at the 5\% statistical significance level.
	}%
    \label{fig:boxplot}
\end{figure}

Finally, Fig.~\ref{fig:lineplot.irmas} shows the evaluation on the IRMAS subset comprising 60 signals.
This evaluation supports the previous analysis and leads to the conclusion that the proposed method is superior to the others. For shorter gaps, DPAI with context is preferable over DPAI not exploiting context.

\begin{figure}[t]
    \centering
    %
    %
    \adjustbox{width=0.72\linewidth}{
%
%
\definecolor{mycolor1}{rgb}{0.00000,0.44700,0.74100}%
\definecolor{mycolor2}{rgb}{0.85000,0.32500,0.09800}%
\definecolor{mycolor4}{rgb}{0.92900,0.69400,0.12500}%
\definecolor{mycolor5}{rgb}{0.49400,0.18400,0.55600}%
\begin{tikzpicture}[trim axis left, trim axis right]

\begin{axis}[%
width=4.00in,
height=2.55in,
at={(0.758in,0.481in)},
scale only axis,
xmin=1,
xmax=6,
xtick={1, 2, 3, 4, 5, 6},
xlabel style={text=white!15!black},
xlabel={gap length},
ymin=0,
ymax=60,
ylabel style={text=white!15!black},
ylabel={SNR (dB)},
axis background/.style={fill=white},
xmajorgrids,
ymajorgrids,
legend style={legend cell align=left, align=left, draw=white!15!black}
]

\addplot[area legend, draw=black, fill=mycolor1, draw opacity=0.1, fill opacity=0.1, forget plot]
table[row sep=crcr] {%
x	y\\
1	22.8362279629937\\
2	17.4966483353934\\
3	13.8991823555901\\
4	12.6984475235494\\
5	10.3922389247237\\
6	9.26830268397424\\
6	10.3773045088493\\
5	11.583577628807\\
4	14.1460195525276\\
3	15.5260594613066\\
2	20.1917660255544\\
1	25.3397590999594\\
}--cycle;
\addplot [color=mycolor1, line width=1.0pt, dashed]
  table[row sep=crcr]{%
1	24.2272879294368\\
2	19.1689259614778\\
3	14.770654763522\\
4	13.375323724723\\
5	10.9560200766075\\
6	9.78619694303379\\
};
\addlegendentry{DPAI without context}

\addplot[area legend, draw=black, fill=mycolor2, draw opacity=0.1, fill opacity=0.1, forget plot]
table[row sep=crcr] {%
x	y\\
1	29.6977663605082\\
2	20.4832501237416\\
3	14.8283360706393\\
4	13.1509534061381\\
5	10.6029166801128\\
6	9.53357021928344\\
6	10.471955672191\\
5	11.7993001654322\\
4	14.5781102234975\\
3	16.2923327430822\\
2	22.2050061651415\\
1	31.9586014226468\\
}--cycle;
\addplot [color=mycolor2, line width=1.0pt, dotted]
  table[row sep=crcr]{%
1	30.9167434477012\\
2	21.3699627281417\\
3	15.5574922407242\\
4	13.7959400912164\\
5	11.1624050280334\\
6	9.97346909879092\\
};
\addlegendentry{DPAI with context}

\addplot[area legend, draw=black, fill=mycolor4, draw opacity=0.1, fill opacity=0.1, forget plot]
table[row sep=crcr] {%
x	y\\
1	11.8738623713574\\
2	10.2012502807043\\
3	8.56964659609479\\
4	8.19757474607603\\
5	6.84479277252158\\
6	6.36635199734184\\
6	7.2526420923118\\
5	7.82621473453551\\
4	9.53078414721063\\
3	9.66224750388923\\
2	11.5911270453548\\
1	13.2665005182421\\
}--cycle;
\addplot [color=mycolor4, line width=1.0pt, dashdotted]
  table[row sep=crcr]{%
1	12.4985507788448\\
2	10.7967471666389\\
3	9.08377400556427\\
4	8.76500933585825\\
5	7.30662060805943\\
6	6.76801872003689\\
};
\addlegendentry{gap-wise Janssen}

\addplot[area legend, draw=black, fill=mycolor5, draw opacity=0.1, fill opacity=0.1, forget plot]
table[row sep=crcr] {%
x	y\\
1	61.1651696553551\\
2	51.2794537750675\\
3	30.1927768983386\\
4	20.0555064275225\\
5	14.6948274553427\\
6	12.2010009777936\\
6	13.2709827083334\\
5	16.2874359888915\\
4	22.0587023080746\\
3	33.2504971278908\\
2	56.0659165705072\\
1	65.4979493623338\\
}--cycle;
\addplot [color=mycolor5, line width=1.0pt]
  table[row sep=crcr]{%
1	63.0738365640512\\
2	53.5730414911606\\
3	31.6821751527086\\
4	21.0062266783803\\
5	15.3841475615997\\
6	12.7243706410715\\
};
\addlegendentry{Janssen-TF-ADMM}

\end{axis}
\end{tikzpicture}%
}
    \adjustbox{width=0.72\linewidth}{
%
%
\definecolor{mycolor1}{rgb}{0.00000,0.44700,0.74100}%
\definecolor{mycolor2}{rgb}{0.85000,0.32500,0.09800}%
\definecolor{mycolor4}{rgb}{0.92900,0.69400,0.12500}%
\definecolor{mycolor5}{rgb}{0.49400,0.18400,0.55600}%
\begin{tikzpicture}[trim axis left, trim axis right]

\begin{axis}[%
width=4.00in,
height=2.55in,
at={(0.758in,0.481in)},
scale only axis,
xmin=1,
xmax=6,
xtick={1, 2, 3, 4, 5, 6},
xlabel style={text=white!15!black},
xlabel={gap length},
ymin=-3.5,
ymax=0,
ylabel style={text=white!15!black},
ylabel={ODG},
axis background/.style={fill=white},
xmajorgrids,
ymajorgrids,
legend style={legend cell align=left, align=left, draw=white!15!black}
]

\addplot[area legend, draw=black, fill=mycolor1, draw opacity=0.1, fill opacity=0.1, forget plot]
table[row sep=crcr] {%
x	y\\
1	-1.87591366666886\\
2	-2.00927185101449\\
3	-2.46818084045233\\
4	-2.72851266923527\\
5	-2.95224242397689\\
6	-3.04889835214943\\
6	-2.97557789657475\\
5	-2.86319296849373\\
4	-2.55588389345397\\
3	-2.21446034799866\\
2	-1.58594653362968\\
1	-1.39415639628966\\
}--cycle;
\addplot [color=mycolor1, line width=1.0pt, dashed]
  table[row sep=crcr]{%
1	-1.64321873796137\\
2	-1.78546664703791\\
3	-2.33913800183789\\
4	-2.65180894161652\\
5	-2.91028493388279\\
6	-3.01685958974133\\
};
\addlegendentry{DPAI without context}

\addplot[area legend, draw=black, fill=mycolor2, draw opacity=0.1, fill opacity=0.1, forget plot]
table[row sep=crcr] {%
x	y\\
1	-0.671523826460564\\
2	-1.39024089943087\\
3	-2.26577234334669\\
4	-2.62105964559212\\
5	-2.91798036131701\\
6	-3.00845321784417\\
6	-2.93631949630714\\
5	-2.83023511923461\\
4	-2.4463822857179\\
3	-1.98077694298249\\
2	-1.02118306741509\\
1	-0.380395225751594\\
}--cycle;
\addplot [color=mycolor2, line width=1.0pt, dotted]
  table[row sep=crcr]{%
1	-0.477132552521142\\
2	-1.19076080965216\\
3	-2.13086836546502\\
4	-2.54054573826156\\
5	-2.87705254534864\\
6	-2.97517950167458\\
};
\addlegendentry{DPAI with context}

\addplot[area legend, draw=black, fill=mycolor4, draw opacity=0.1, fill opacity=0.1, forget plot]
table[row sep=crcr] {%
x	y\\
1	-3.12536301977579\\
2	-3.19939342509814\\
3	-3.23740608991054\\
4	-3.25128700264256\\
5	-3.27430546352769\\
6	-3.2823591508702\\
6	-3.24612286335337\\
5	-3.22742919106901\\
4	-3.20665570027035\\
3	-3.18416686508191\\
2	-3.12937517912897\\
1	-3.03594806175738\\
}--cycle;
\addplot [color=mycolor4, line width=1.0pt, dashdotted]
  table[row sep=crcr]{%
1	-3.09217155210078\\
2	-3.17368862416697\\
3	-3.21828645361849\\
4	-3.23557501681071\\
5	-3.25720943211263\\
6	-3.26755020320695\\
};
\addlegendentry{gap-wise Janssen}

\addplot[area legend, draw=black, fill=mycolor5, draw opacity=0.1, fill opacity=0.1, forget plot]
table[row sep=crcr] {%
x	y\\
1	-0.0597380087864835\\
2	-0.31191075591863\\
3	-1.48331689293575\\
4	-2.607962243211\\
5	-2.99488341390913\\
6	-3.13003450131689\\
6	-3.0626048821919\\
5	-2.88561425431654\\
4	-2.38885810163165\\
3	-1.21224804695806\\
2	-0.16962614238671\\
1	-0.0260704890850144\\
}--cycle;
\addplot [color=mycolor5, line width=1.0pt]
  table[row sep=crcr]{%
1	-0.0385683580675432\\
2	-0.22344657702772\\
3	-1.35272685527161\\
4	-2.51967682401608\\
5	-2.95401543592271\\
6	-3.10571366146828\\
};
\addlegendentry{Janssen-TF-ADMM}

\end{axis}
\end{tikzpicture}%
}%
    %
    %
    %
    \vspace{-7pt}%
    \caption{%
        SNR and ODG results on the IRMAS dataset.
    }
    \label{fig:lineplot.irmas}
\end{figure}

Regarding computational demands,
a~single run of DPAI takes ca 19 minutes on the NVIDIA Tesla V100S GPU with 32\,GB of memory, regardless of the number of missing spectrogram columns.
Reconstructing a~single signal with Janssen-TF-ADMM
takes 10 to 20~minutes,
proportional to the number of columns;
this performance corresponds to a~PC with Intel Core i7 3.40\,GHz processor, 32\,GB RAM. 


\section{Conclusion and Outlook}
The paper has shown that the proposed method of spectrogram inpainting, Janssen-TF,
performs significantly better than the recently introduced DPAI algorithm, which is based on the deep prior idea.
The conclusion has been certified both by objective and by subjective tests. 
%
In the future, it might be interesting to use the same adaptation strategy with other methods for the time-domain inpainting,
such as the PHAIN regularized algorithm
\cite{TanakaYatabeOikawa2024:PHAIN}.


Matlab
source
codes for Janssen-TF and other supplementary material are
publicly
available at
\url{https://github.com/rajmic/spectrogram-inpainting}.


\vspace*{3ex}
\newcommand{\noopsort}[1]{} \newcommand{\printfirst}[2]{#1} \newcommand{\singleletter}[1]{#1} \newcommand{\switchargs}[2]{#2#1}

\end{document}